\documentclass{elsart}

\usepackage{natbib}

 \usepackage{epsfig}

\usepackage{amssymb}

\begin{document}

\begin{frontmatter}

% Title, authors and addresses

% use the thanksref command within \title, \author or \address for footnotes;
% use the corauthref command within \author for corresponding author footnotes;
% use the ead command for the email address,
% and the form \ead[url] for the home page:
% \title{Title\thanksref{label1}}
% \thanks[label1]{}
% \author{Name\corauthref{cor1}\thanksref{label2}}
% \ead{email address}
% \ead[url]{home page}
% \thanks[label2]{}
% \corauth[cor1]{}
% \address{Address\thanksref{label3}}
% \thanks[label3]{}

\title{Observational studies of stellar black hole binaries and ULXs}

% use optional labels to link authors explicitly to addresses:
% \author[label1,label2]{}
% \address[label1]{}
% \address[label2]{}

\author[a1]{Aya Kubota},
\author[a1,a2]{Kazuo Makishima}   

\address[a1]{Institute of Physical and Chemical Research, \\
2-1 Hirosawa, Wako-shi, Saitama 351-0198, Japan}
\address[a2]{Department of Physics,  University of Tokyo,\\
7-3-1 Hongo, Bunkyo-ku, Tokyo 113-0033, Japan}

\begin{abstract}
% Text of abstract
We outline a framework for understanding the
X-ray spectra of high mass accretion rate
stellar black holes based on X-ray data from RXTE and ASCA. 
Three spectral regimes can be separated out by the behaviour of the 
observed disk luminosity and temperature.  
The well established "standard regime" is seen when the disk dominates the
spectrum, where only a small fraction of the luminosity is emitted in the power law
tail. These spectra generally 
satisfy the standard relation expected for thermal emission from a constant
area, namely that the 
disk bolometric luminosity, $L_{\rm disk}$, is proportional to 
its maximum temperature, $T_{\rm in}^4$. However, at higher luminosities
this starts to change to $\propto T_{\rm in}^2$. This 
"apparently standard regime" is still dominated by the disk emission, but 
this difference luminosity-temperature relation and subtle changes 
in spectral shape may show that 
another cooling process is required in addition to radiative cooling.
At intermediate luminosities there is an anomalous regime (or  weak very high state) 
where the disk temperature and luminosity are less clearly related. These 
spectra are characterized by the presence of a much stronger comptonized tail 
indicating high energy electrons. When observed disk emission is 
corrected for the the effects of comptonisation then these points lie back on the 
standard relation. The growth of this comptonising corona is also clearly linked to 
the quasi-periodic oscillations, as these are
observed preferentially in the anomalous regime. This presented picture was
found to explain the spectral behavior of both black hole binaries in our
Galaxy and LMC. Spectral evolution of several bright ULXs observed with ASCA
were also successfully explained in the same picture.  

\end{abstract}


\end{frontmatter}

\section{Classical understanding of stellar black holes in high accretion rate state}

\begin{figure}
\begin{center}
\includegraphics[clip=true,width=0.45\textwidth,angle=-90]
{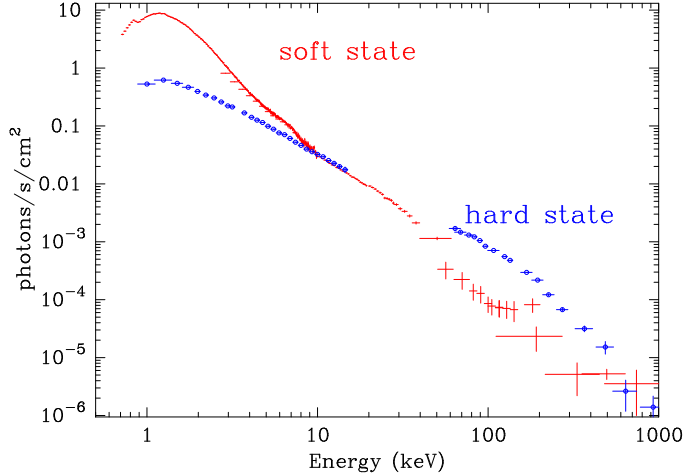}%fig1
\end{center}
\caption{X/$\gamma$-ray spectra of Cyg X-1 in both the high/soft and the low/hard state, taken from Cui et al. (1998), Gierlinski et al. (1997), Balucinska \& Hasinger (1991), Phlips et al. (1996), and McConnell et al. (1994).}
\label{fig:cygx1}
\end{figure}

In a close binary consisting of a stellar-mass
black hole and a mass donating normal star,
the accreting matter %from the star 
releases its gravitational
energy as X-ray radiation.
In Fig.~\ref{fig:cygx1},
we show two typical X-ray spectra of Cyg X-1, the 
best studied black hole binary, showing its two distinct 
spectral states. 
In the low/hard state, its spectrum is approximately a power-law,
with an exponential cut-off near $\sim100$ keV. 
By contrast, the high/soft state is characterized 
by dominant soft emission below 10 keV,
accompanied by a hard power-law tail extending to 500 keV and beyond.
%When the mass accretion rate $\dot{M}$ is high, 
%such a black hole binary is usually found in a
%so-called high/soft state, of which X-ray spectrum is
%characterized by a very soft component accompanied by a
%power-law tail.

As described, e.g., by Makishima et al. (1986),
the soft spectral component is 
interpreted as thermal emission
from an optically thick accretion disk around the black hole, 
as it can be well reproduced by
a multi-color disk model (MCD model; Mitsuda et al. 1984).
This model approximates a spectrum 
from a standard accretion disk (Shakura \& Sunyaev 1973).
Under this standard disk solution, 
the Keplerian rotational energy takes up half of the gravitational energy,
while the remaining half is radiated as approximately blackbody emission from 
the disk surface. The MCD model ignores any inner disc boundary condition, 
describing the disk local temperature $T(r)$ as 
\begin{equation}
T(r)=T_{\rm in}\cdot \left(\frac{r}{r_{\rm in}}\right)^{-3/4}~~,
\label{eq:tr}
\end{equation}
where $T_{\rm in}$ and $r_{\rm in}$ are the two spectral parameters of the MCD model,
the maximum observed disk temperature and the apparent disk inner radius, respectively.
The latter is related to the true inner radius, $R_{\rm in}$ by
\begin{equation}
R_{\rm in}=\kappa ^2 \cdot \xi \cdot r_{\rm in}~~,
\label{eq:rin}
\end{equation}
where $\xi=0.41$ is a correction to account for a stress free inner boundary condition 
(Kubota et al.~1998) and $\kappa\sim1.7$--2.0 is a colour temperature correction
factor to account for distortions of the spectrum from a true blackbody
(Shimura \& Takahara 1995).
The disk bolometric luminosity $L_{\rm disk}$ can be related to these two spectral 
parameters as 
\begin{equation}
L_{\rm disk}=4\pi r_{\rm in}^2 \sigma T_{\rm in}^4~~,
\label{eq:ldisk}
\end{equation}
where $\sigma$ is the Stefan-Boltzmann constant.
This model function is simple, and well reproduces the soft 
spectral component of high/soft state black hole binaries with 
physically meaningful fit parameters.
%The values of $r_{\rm in}$ were found to be 
%kept constant as the disk luminosity changes significantly, 
%and the values of $R_{\rm in}$ estimated via equation (1.2) are consistent with the 
%innermost Keplerian orbit, $R_{\rm ms}$, for a central black hole.
For example, the soft component of Cyg X-1 shown in Fig.~\ref{fig:cygx1} 
was well fitted with the MCD model of $R_{\rm in}=90$~km 
(Dotani et al. 1997) with $\kappa=1.7$ and $\xi=0.41$. 
Since the black hole mass of Cyg X-1 was estimated to be 10--16~$M_\odot$ 
from several optical observations
(Gies \& Bolton 1986, Ninkov et al. 1987, Liang \& Nolan 1983), 
the last Keplarian orbit, $R_{\rm ms}$, of a non-spinning black hole is calculated as 
$R_{\rm ms}=6R_{\rm g}(=6GM/c^2)=89$--140~km for Cyg~X-1. 
Therefore, it is found that $R_{\rm in}$ determined by the X-ray observations 
well agrees with $R_{\rm ms}$ calculated by referring to the optical observations.

\begin{figure}
\begin{center}
\includegraphics[clip=true,width=0.45\textwidth,angle=-90]
{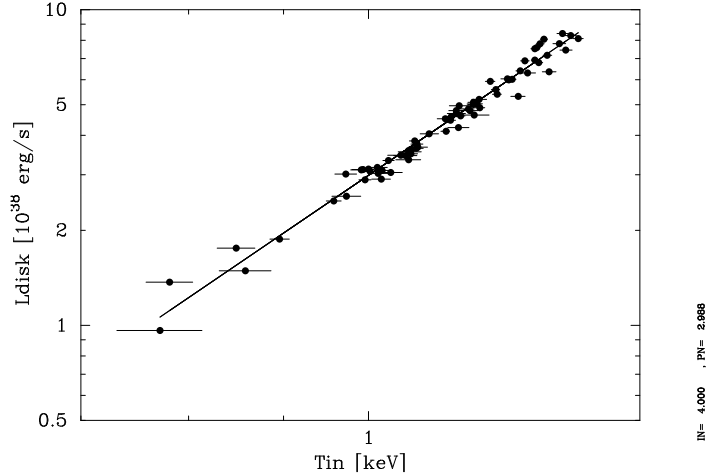}%fig2
\end{center}
\caption{The estimated $L_{\rm disk}$ of LMC X-3 are plotted against the observed 
$T_{\rm in}$. The source distance and inclination are assumed to be $D=50$~kpc 
and $i=66^\circ$. The solid line indicates $L_{\rm disk}\propto T_{\rm in}^4$.}
\label{fig:lmcx3}
\end{figure}

Even more compelling evidence for the standard disk formalism is 
given by the observational fact that 
the value of $R_{\rm in}$ is usually observed to 
remain constant in the high/soft state
as $L_{\rm disk}$ changes significantly. 
This indicates that 
$R_{\rm in}$ reflects the physical parameters of the black hole itself. 
Figure~\ref{fig:lmcx3} shows the values of $L_{\rm disk}$ obtained by 
spectral fits of $\sim$80 RXTE/PCA observations of high/soft state spectra from 
LMC~X-3, a well known black hole binary system. 
This figure shows that all the data points lie along the solid line which 
represents $L_{\rm disk}\propto T_{\rm in}^4$ (requiring constant 
emitting area i.e. constant $r_{\rm in}$, and constant colour temperature correction,
$\kappa$, in equation~\ref{eq:ldisk}) over the range of 
$L_{\rm disk}=1$--8$\times 10^{38}~{\rm erg~s^{-1}}$($\sim$10--80\% of the 
Eddington limit, $L_{\rm E}$, of a $6~M_\odot$ black hole.

This relation of $L_{\rm disk}\propto T_{\rm in}^4$ has been widely observed 
in many high/soft-state black hole binaries except at 
very high disk temperature (typically at $>0.9$--1.2~keV, see below). 
%Many of the high/soft-state black hole binaries have thus been found to be well 
%explained by the standard picture
(e.g., Tanaka \& Lewin 1995; McClintock \& Remillard 2003; Ebisawa et al. 1994; 
Gierlinski \& Done 2004). 

Despite these sucesses, standard disc models strongly conflict
with the observed disc emission on several points. 
(1) The $\alpha$ viscosity parameterisation of the standard 
Shakura-Sunyaev disc model is unstable in the range $\sim$ 0.1--1 $L_{\rm E}$
where radiation pressure is dominant but the disc is not yet optically thick enough to
regain stability through advecting the radiation 
(slim disk solution; Abramowicz et al. 1988). Yet the observed 
disk emission is very stable in this luminosity range.
(2) Theoretical models of radiation transfer through the disc vertical predict that 
the colour temperature correction factor should change with luminosity.
However, the amount of this change has now been shown to be rather
small when true metal opacities are included (Davis et al 2005).
This superceeds previous work which included only 
H and He, where the calculations of 
Merloni et al (2000) predicting a large change in $\kappa$, while
those of Shimura \& Takahara (1995) had
$\kappa$ varying by less than 20\%. 
The resolution of this discrepancy may be that
Merloni et al (2000) assumed a vertical density profile which is appropriate 
for a radiation pressure dominated disc, but their lowest luminosity discs are
gas pressure dominated (Gierlinski \& Done 2004). 

%Observationally, the high/soft state has a {\em stable} disc with approximately 
%{\em constant} colour temperature correction over
% wide range of $L_{\rm disk}$ as shown in Fig.~\ref{fig:lmcx3}. 
%including a slim disk solution, which takes advective cooling
%into account (Abramowicz et al. 1988; Watarai et al. 2000).
Thus while the second point is now not necessarily in conflict with the standard
disk models, the stability of the observed high/soft state spectra 
can give constraints on theoretical models of accretion disc structures. 

\section{Breakdown of the standard disk}

\begin{figure}[htbp]
%fig3a fig3b
\begin{center}
\begin{minipage}{0.45\textwidth}
{\footnotesize (a)GRO J1655--40}\\
\includegraphics[clip=true,width=4.5cm,angle=-90]
{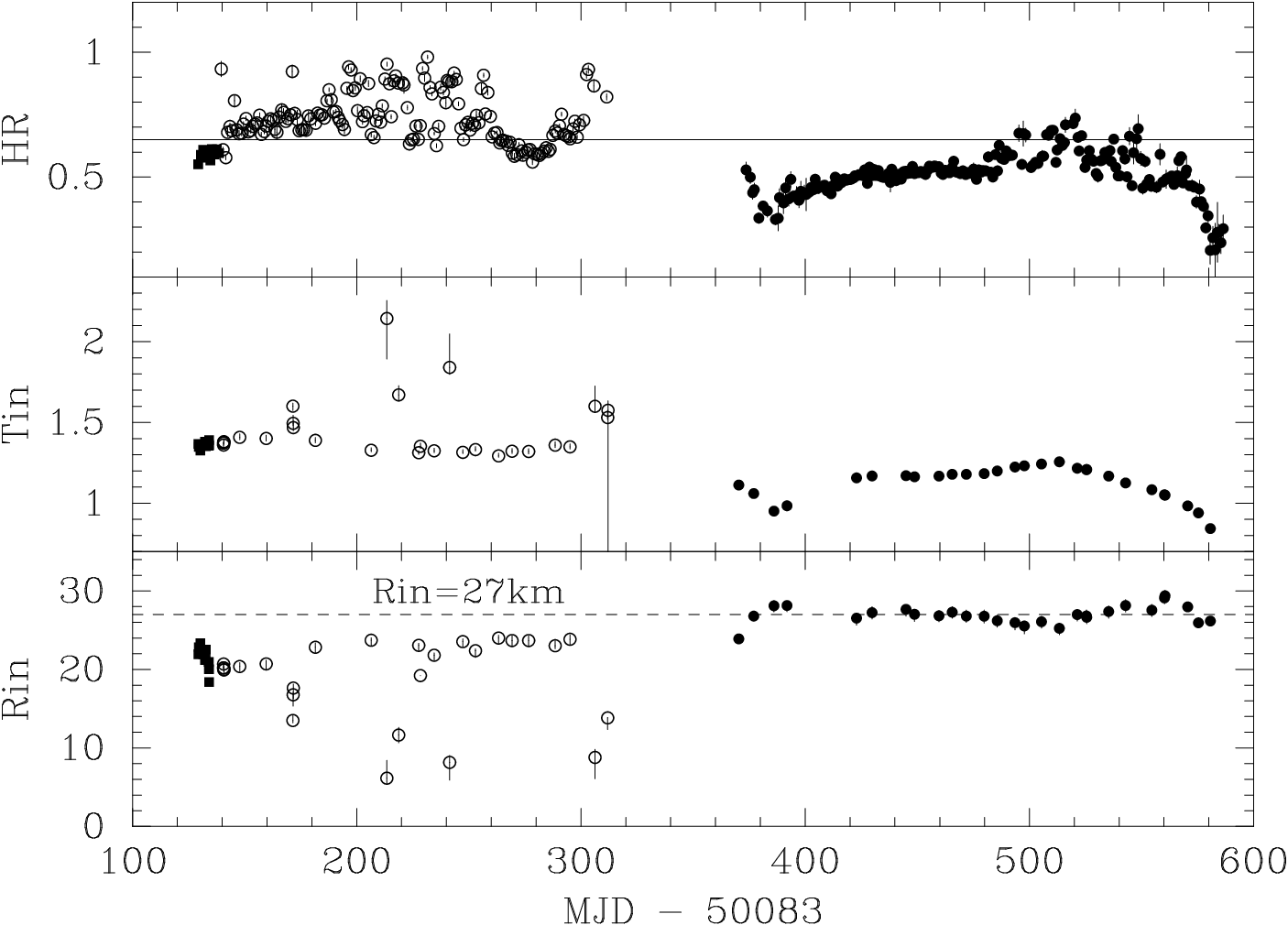}
\end{minipage}
\begin{minipage}{0.45\textwidth}
{\footnotesize (b)XTE J1550-564}\\
\includegraphics[clip=true,width=4.5cm,angle=-90]
{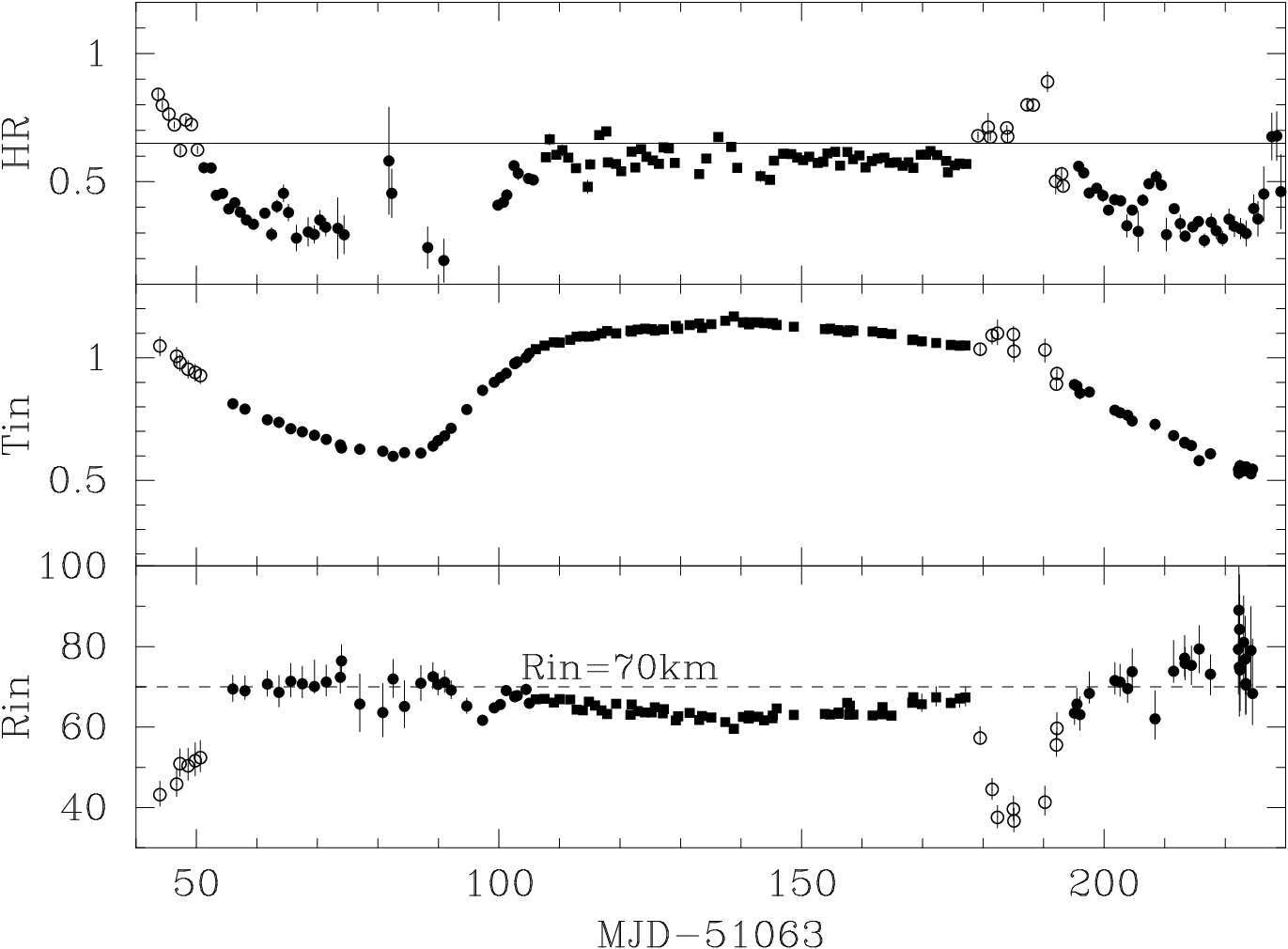}
\end{minipage}
\end{center}
\caption{Hardness ratio (5--12keV/3-5keV) by the RXTE/ASM and values of 
$T_{\rm in}$ and $R_{\rm in}$ obtained by the MCD plus power-law fit to the RXTE/PCA data of 
GRO J$1655-40$ (a) and XTE J$1550-564$ (b). 
Filled circle, open circle, and filled square indicate 
the standard regime (standard disk), 
the anomalous regime (very high state; strong comptonization), and the 
apparently standard regime (slim disk), respectively. 
These are taken after KME01 and KM04.}
\label{fig:hr}
\end{figure}

However, high $L/L_{Edd}$ spectra are not always disk dominanted.
There is an alternative bright spectral state, termed "very high" or "steep power law" 
state which was discovered by Miyamoto et al. (1991) and is 
characterised by both strong disk {\em and} strong (and rather steep) 
power law emission which is highly variable
(e.g. van der Klis 1994; McClintock \& Remillard 2004).
When the spectra in this state are fitted with the canonical MCD plus 
power-law model, the data no longer show $L\propto T^4$, 
implying that the estimated values of $R_{\rm in}$ (and/or the
colour temperature correction, $\kappa$) are no longer constant. 
This anomalous behavior is clearly seen in Fig.~\ref{fig:hr} 
which shows results of the MCD plus power-law fit to the
PCA data of two high state black hole binaries, 
GRO~J$1655-40$ and XTE~J$1550-564$
(Kubota, Makishima, Ebisawa 2001, KME01; Kubota \& Makishima 2004, KM04).
The data points indicated with open circles show significant variation in both 
$T_{\rm in}$ and $R_{\rm in}$, while those with filled circles show
constant values of $R_{\rm in}$. 
This anomaly was found when the spectra have a strong power law tail, carrying more
than $\sim$ 30 per cent of the luminosity
(corresponding to a hardness ratio larger than $\sim$~0.6--0.7 shown in the
top panels of Fig.~\ref{fig:hr}).

Another breakdown in the $L_{disk}\propto T^4$ relation can 
also be seen in Fig.~\ref{fig:hr} for highly luminous 
{\em disk dominated} spectra with 
negligible power-law tail (filled squares: details are in KME01 and KM04). 
However, the obtained values of $R_{\rm in}$ are slightly 
(but systematically; $\sim$15--25\% in GRO~J$1655-40$ and $\sim$10--15\% in 
XTE~J$1550-564$) 
smaller than those in the 
typical high/soft state indicated with filled circles in Fig.~\ref{fig:hr}.

\begin{figure}[htbp]
%fig4a fig4b
\begin{center}
\begin{minipage}{0.45\textwidth}
{\footnotesize (a)GRO J1655--40}\\

\vspace{-8mm}
\includegraphics[clip=true,width=4.5cm,angle=-90]
{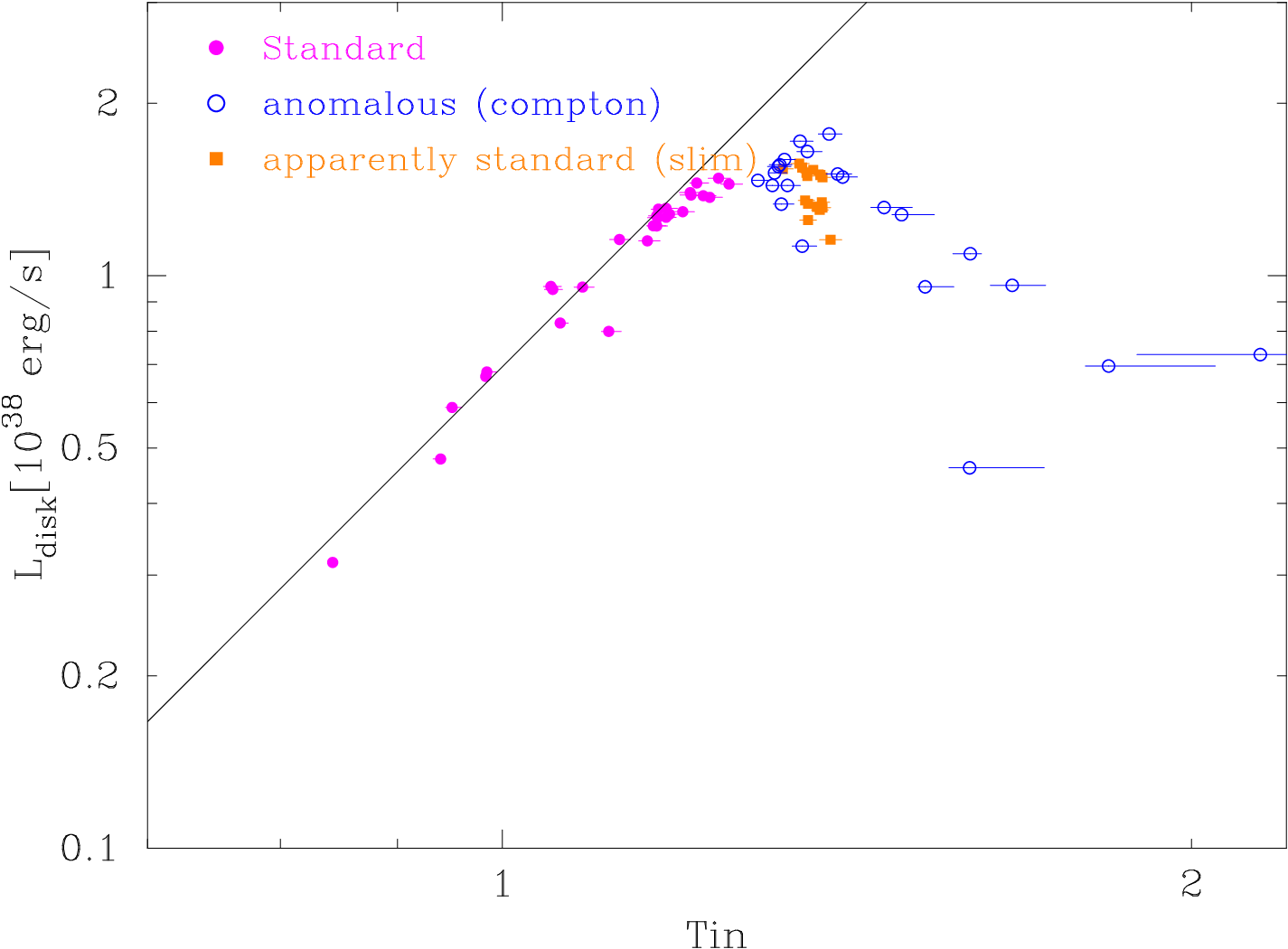}
\end{minipage}
\begin{minipage}{0.45\textwidth}
{\footnotesize (b)XTE J1550-564}\\

\vspace{-8mm}
\includegraphics[clip=true,width=4.5cm,angle=-90]
{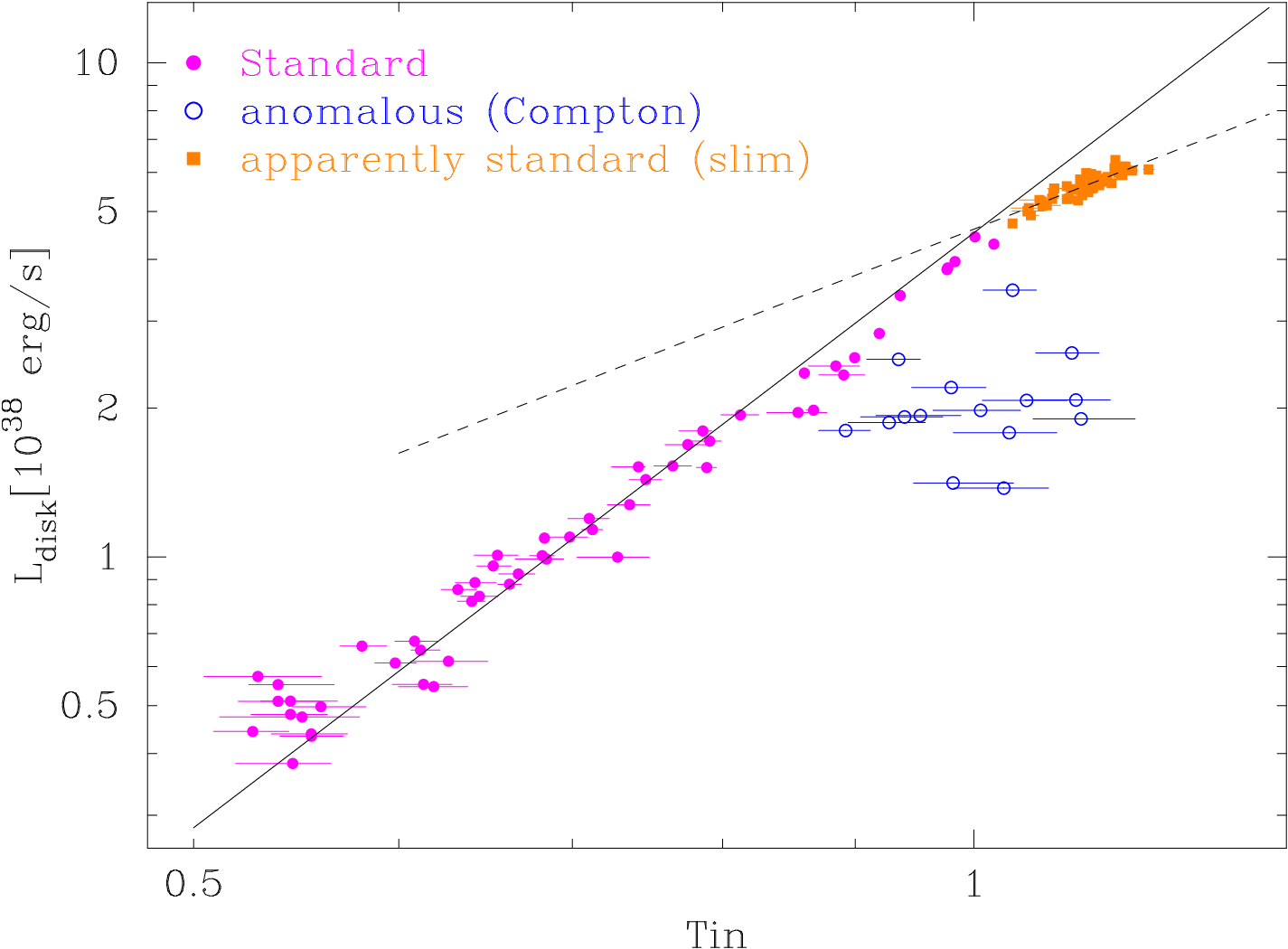}
\end{minipage}
\caption{$L_{\rm disk}$ is plotted against $T_{\rm in}$ based on the 
MCD plus power-law fit (after KME01 and KM04) }
\label{fig:tl1}
\end{center}
\end{figure}

\begin{figure}[htbp]
%fig5a fig5b fig5c
\begin{center}
\begin{minipage}{0.32\textwidth}
\includegraphics[clip=true,width=3cm,angle=-90]
{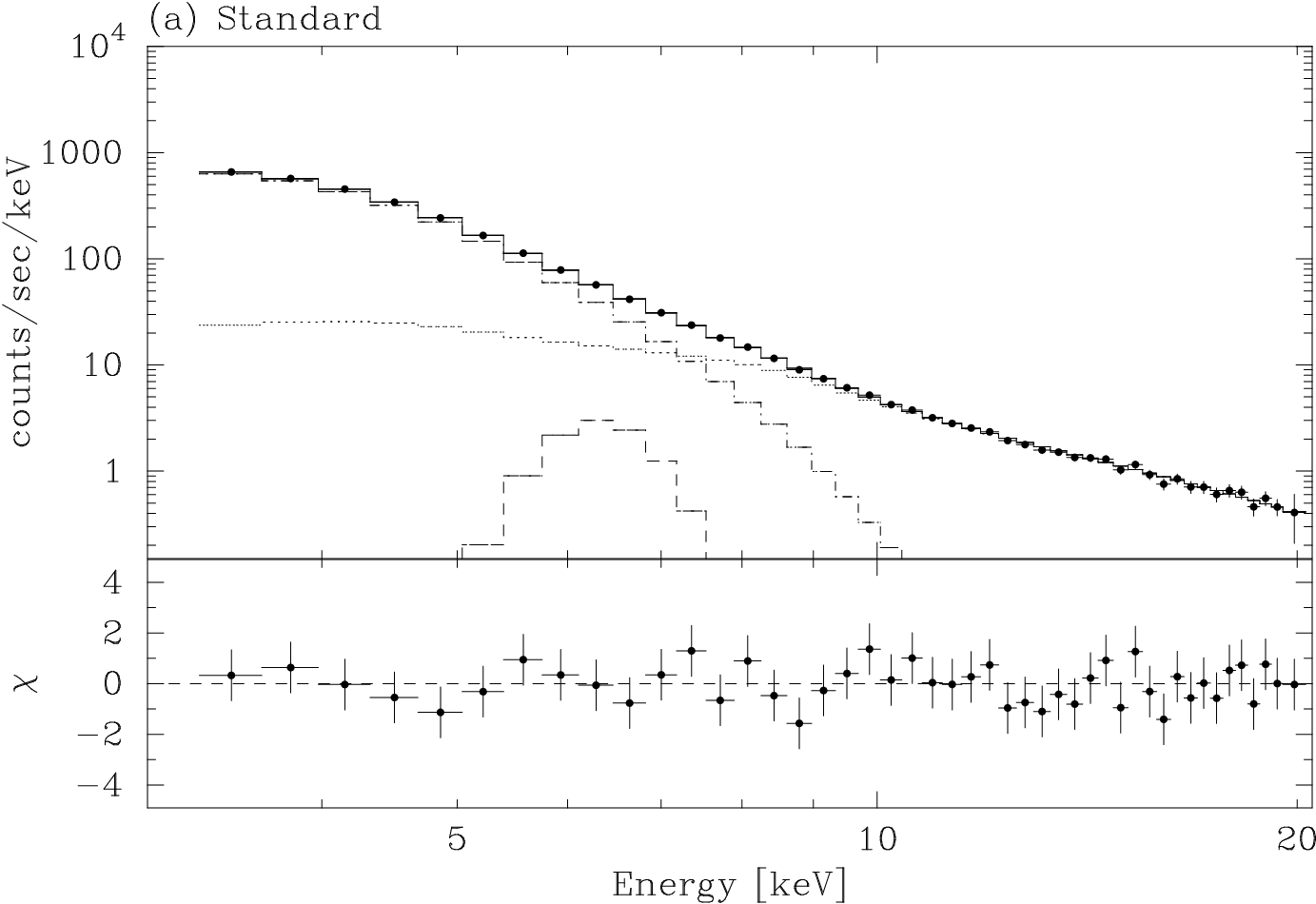}
\end{minipage}
\begin{minipage}{0.32\textwidth}
\includegraphics[clip=true,width=3cm,angle=-90]
{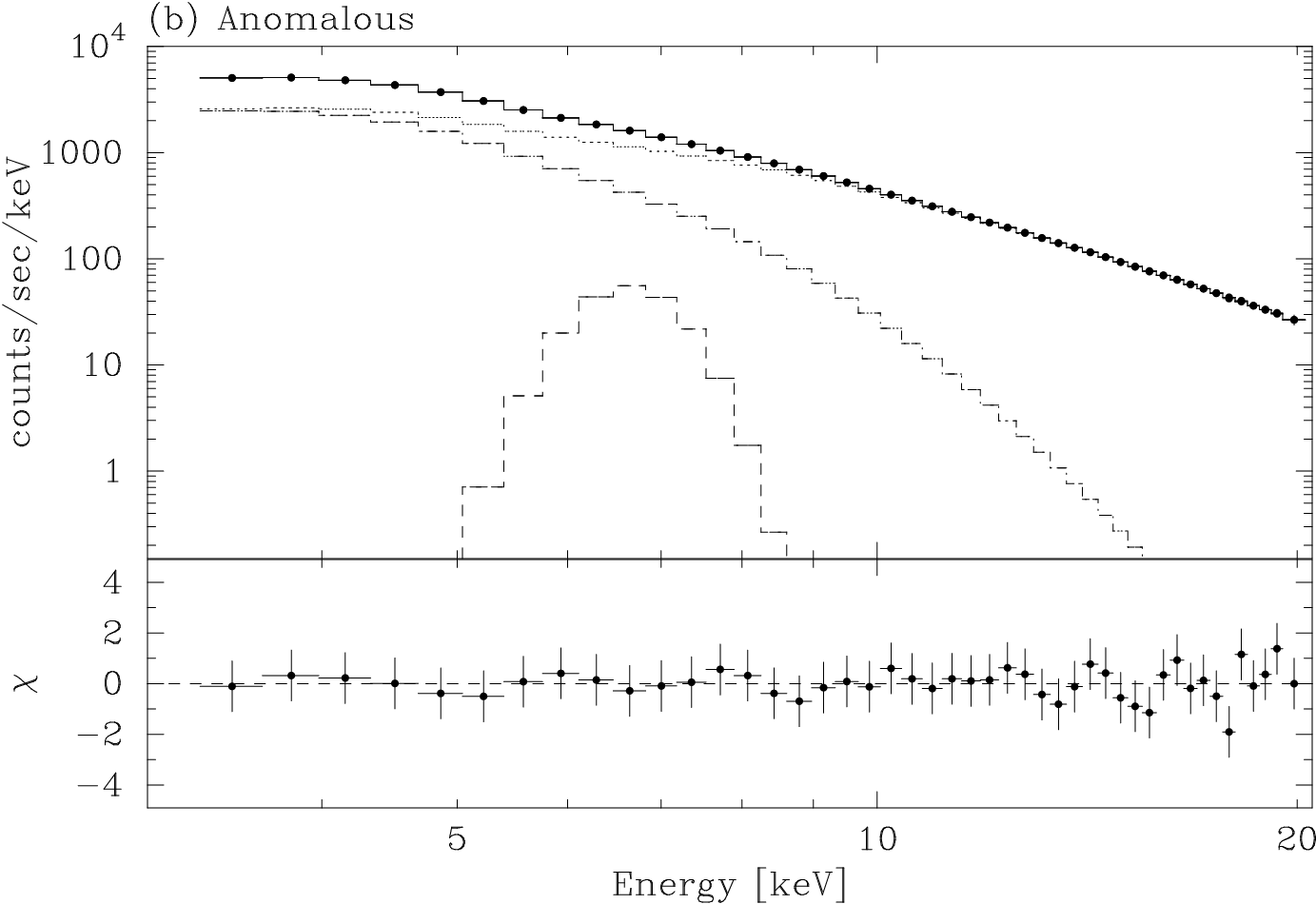}
\end{minipage}
\begin{minipage}{0.32\textwidth}
\includegraphics[clip=true,width=3cm,angle=-90]
{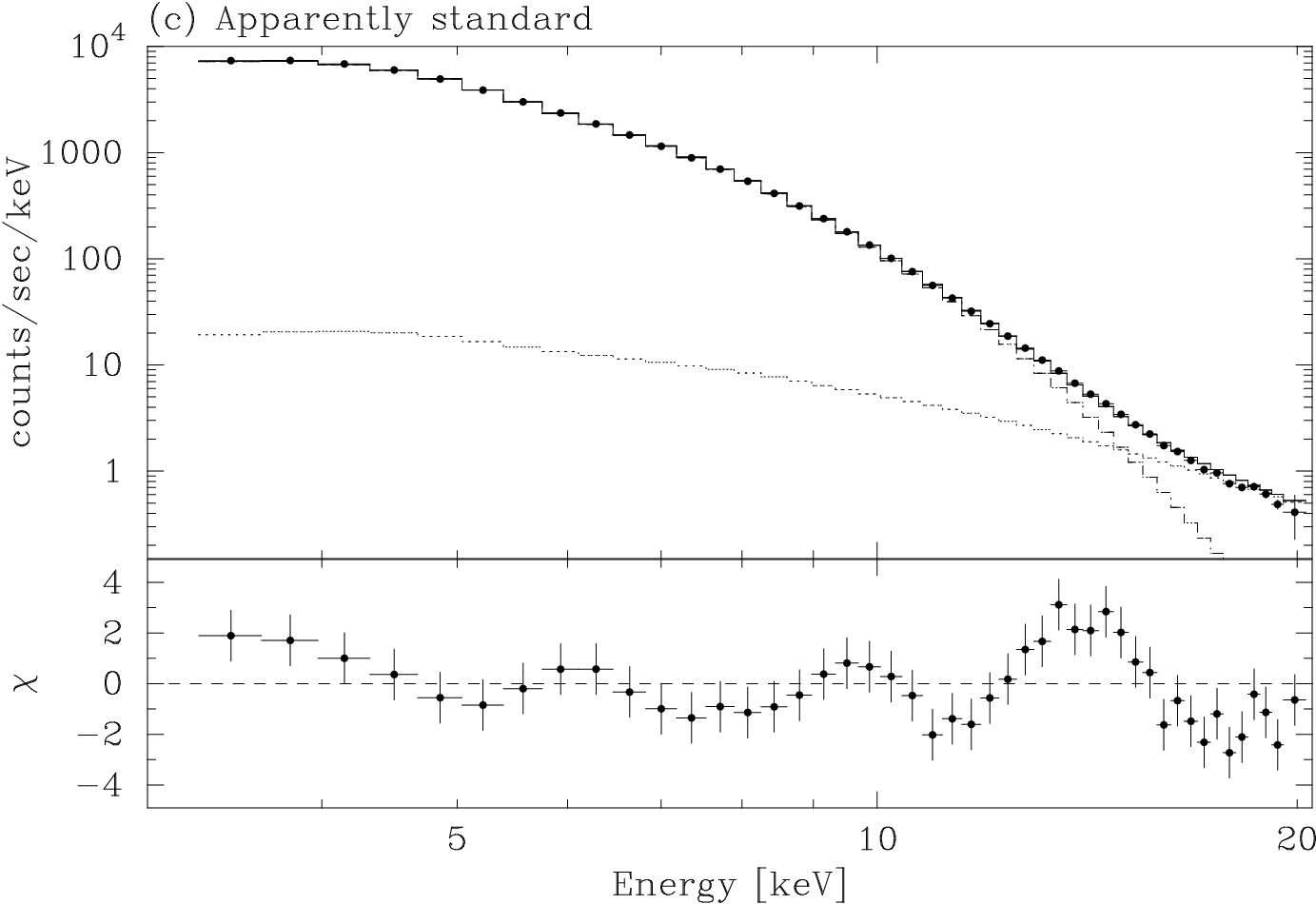}
\end{minipage}
\caption{Typical PCA spectra of XTE~J$1550-564$ fitted with the MCD plus power-law model. The results of the standard regime (a), 
anomalous regime (b) and apparently standard regime (c) are shown after KM04. Note
how the power law contribution is small in the standard and apparently standard regimes,
but is much larger in the anomolous spectrum.}
\label{fig:spec}
\end{center}
\end{figure}

These violations can be characterized in Fig.~\ref{fig:tl1} in which  
$L_{\rm disk}$ is plotted against $T_{\rm in}$. 
In this figure, data points with filled circle are consistent with 
solid lines showing constant  $R_{\rm in}$ via equation~(\ref{eq:ldisk}), and thus they
are in good agreement with the standard picture as is the case of LMC X-3. 
We thus called these data points standard regime in KME01 and KM04. 
On the other hand, the data points with open circles, 
show significant deviation from the 
solid lines in both GRO~J$1655-40$ and XTE~J$1550-564$, so were
termed anomolous regime by KME01 and KM04, or "weak very high state" by 
Kubota \& Done (2004).
The data points with filled squares show the highest $L_{\rm disk}$. 
Their deviation is much milder than that of the anomalous regime, 
and their spectral shape is similar to that of the standard regime.
This was termed "apparently standard regime" by KM04.
Figure~\ref{fig:spec} shows examples of typical PCA spectra of XTE~J$1550-564$ 
in each regime. 

%\vspace{5mm}

%\vspace*{5mm}

%\vspace{5mm}
\begin{figure}[htbp]
%fig6a fig6b
\begin{center}
\begin{minipage}{0.45\textwidth}
{\footnotesize (a)GRO J1655--40 }\\

\vspace{-8mm}
\includegraphics[clip=true,width=4.5cm,angle=-90]
{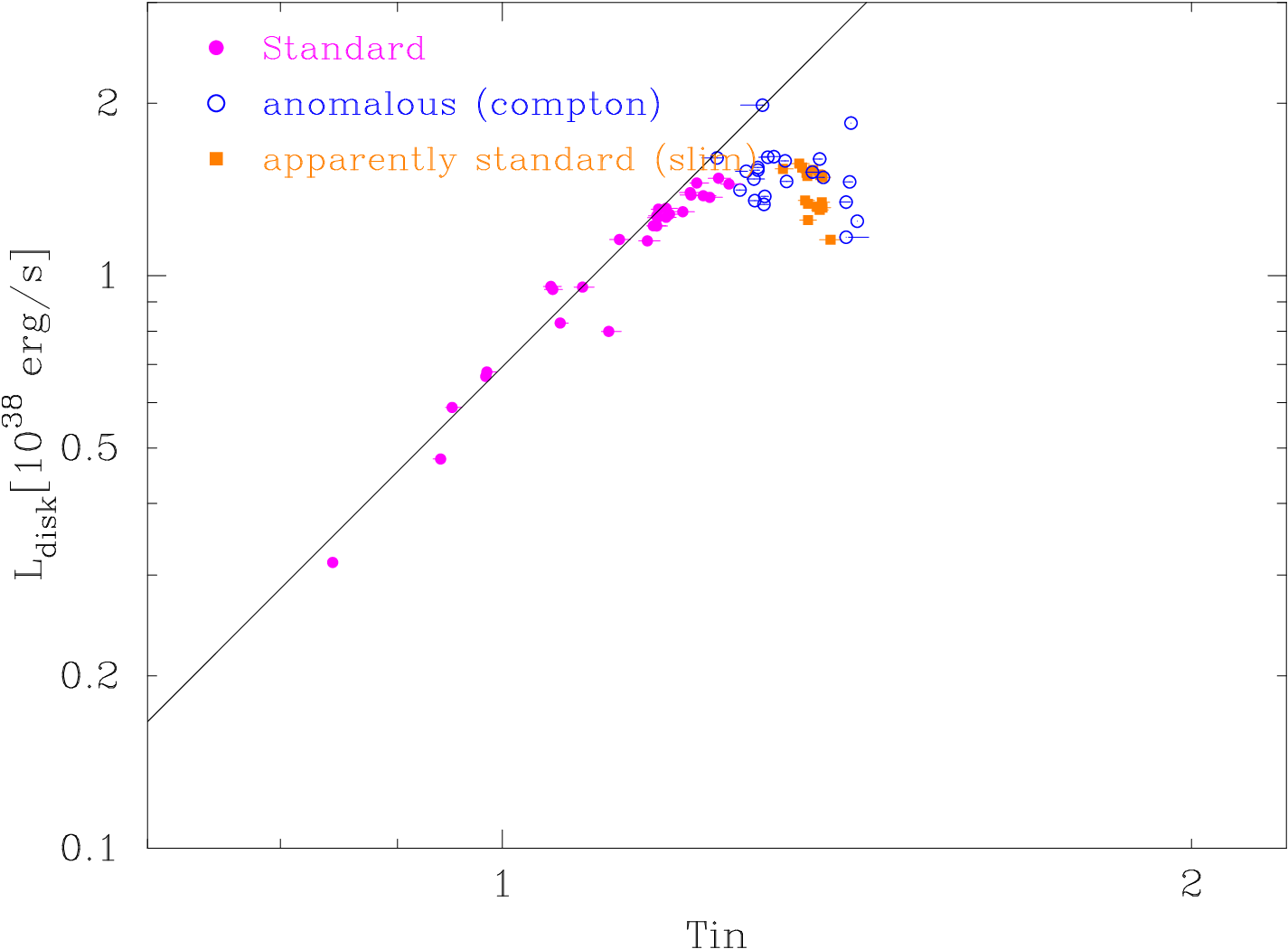}
\end{minipage}
\begin{minipage}{0.45\textwidth}
{\footnotesize (b)XTE J1550-564}\\

\vspace{-8mm}
\includegraphics[clip=true,width=4.5cm,angle=-90]
{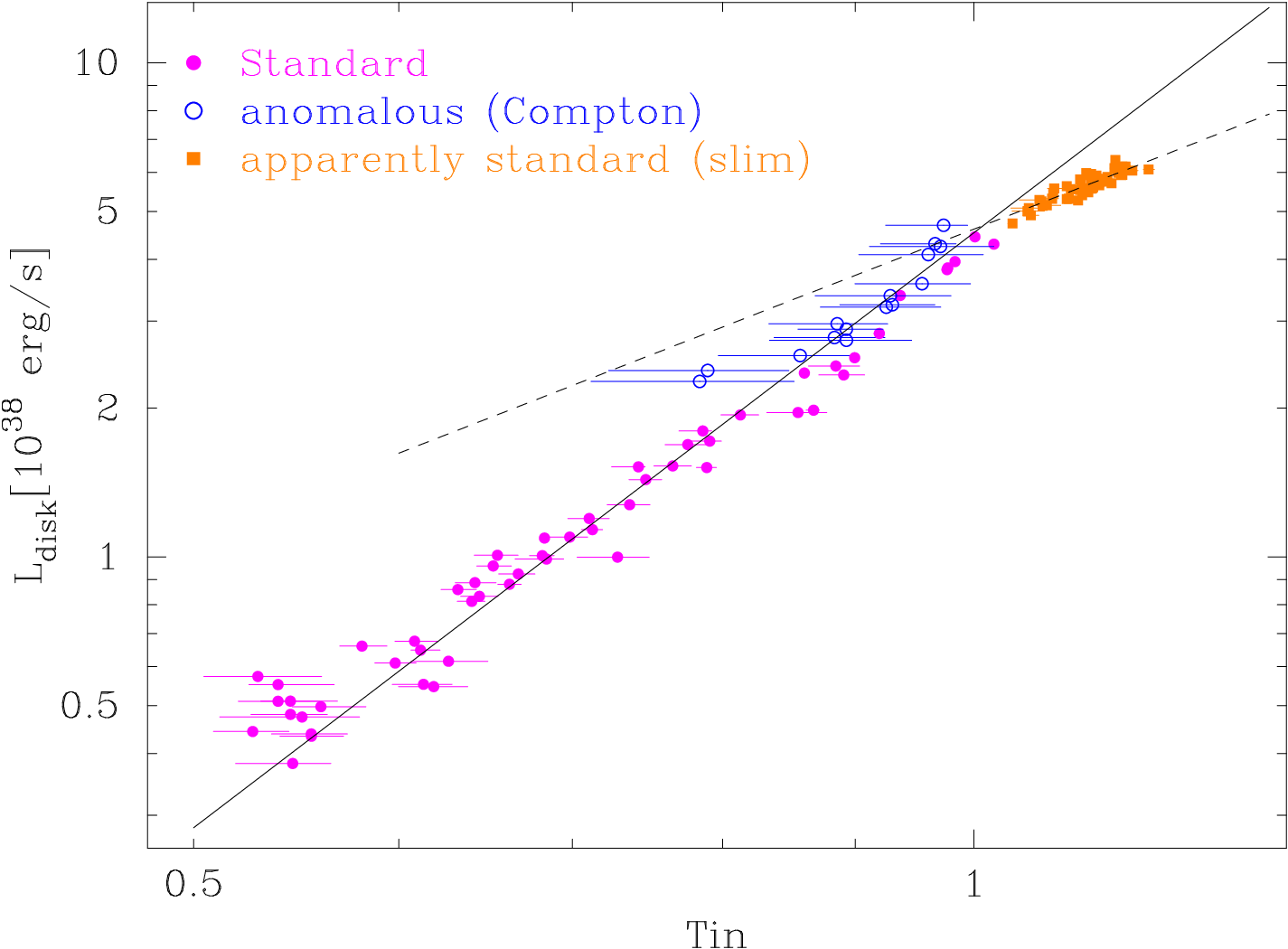}
\end{minipage}
\caption{$L_{\rm disk}+L_{\rm comp}$ is plotted against $T_{\rm in}$ 
based on MCD plus power-law plus Compton component fit (after
KME01 and KM04). }
\label{fig:tl2}
\end{center}
\end{figure}

\section{Understanding the observed problems}

\subsection{Inverse Compton scattering in the anomalous (weak very high) regime}

A clue to the problem in the anomalous regime was given by KME01 and KM04. 
They showed that the anomolous regime was seen when there was 
enhanced hard X-ray power law emission 
compared to the disc dominated spectra. This shows that there is 
significant inverse Compton scattering 
of the disk photons by some high energy electrons, probably in a
corona over the disk. The spectral shape requires that 
the Compton region is mildly optically thick, so the 
disk luminosity is artificially suppressed by $\sim e^{-\tau}$.
In Fig.~\ref{fig:tl2} we approximately correct for this by 
plotting $L_{\rm disk}+L_{\rm thc}$ ($L_{\rm thc}$ is the
bolometric luminosity of the thermal compton component) 
instead of $L_{\rm disk}$, 
against $T_{\rm in}$ based on this spectral fits.
We found that the anomolous regime 
points in  Fig.~\ref{fig:tl1} settled back to the solid line, 
consistent with a constant inner radius disk in this and the disk
dominated states. Other approximations for the
correction to the observed disk luminosity (e.g. assuming that
Compton scattering conserves photon number) give similar results (KM04).

%A low temperature, mildly optically thick region is generically required 
A mildly optically thick, thermal inverse Compton scattering is generically required
in all broad bandpass data of the very high state
(GS 2000+25, GS 1124-68: Zycki, Done \& Smith 1999;
XTE J1915+105: Zdziarski et al 2002; XTE J1550--564, 
Gierlinski \& Done 2003; GRO~J$1655-40$:
Kobayashi et al.~2003).

It is interesting to note here that QPOs are preferentially observed in the 
anomalous regime in both GRO~J$1655-40$ (Remillard et al. 1999) 
and XTE~J$1550-564$ (Remillard et al. 2002), and the energy spectra 
of the QPO are also follow the spectrum of the Comptonised emission.  
Therefore, the QPOs should have some relation to the existence of the Compton cloud.

\subsection{Another cooling process in the apparently standard regime}

The apparently standard regime
corresponds to the most luminous phase of the outburst, and 
the data in this regime occupies the upper-right region of the
$L_{\rm disk}$-$T_{\rm in}$ diagram (Fig.~\ref{fig:tl2}).
%, and
%In this regime, 
%$T_{\rm in}$ gradually changed
%keeping $L_{\rm disk}$ almost constant at 
%$6\times10^{38}\cdot D^2_5~{\rm erg~s^{-1}}$ 
%($\sim0.4~L_{\rm E}$).
%deviates moderately from the
%standard $L_{\rm disk}\propto T_{\rm in}^4$ relation.
%, as $L_{\rm disk}\propto T_{\rm in}^{\sim2}$. 
%The obtained values of $r_{\rm in}$ are not constant but 
%become smaller than those in the standard regime.
%, exhibiting a weak correlation with 
%$T_{\rm in}$ as $r_{\rm in}\propto {T_{\rm in}}^{\sim -1}$ (Fig.~\ref{scat}a).
%
As seen in Fig.~\ref{fig:spec}, 
the spectra in the apparently standard regime
show a dominant soft component accompanied by a very weak hard tail.
Although these properties are similar to the standard regime, 
the apparently standard regime is slightly different from the 
standard regime because the disk luminosity does not increase with $T^4$,
but rather more slowly, more like $T^2$. Thus either the disk inner radius
is decreasing, or the colour temperature correction is increasing
(or both).

%(or moderate saturation of $L_{\rm disk}$).
%and the absence of the Fe-K line feature that is usually found in the 
%{\it standard regime}.

Moreover, KM04 showed that the canonical spectral 
model often failed to give acceptable fits to the data 
(Fig.~\ref{fig:spec}; see also Fig.~2 in KM04). 
In order to quantify the difference, KM04 constructed 
a generalization of the MCD model in which the radial 
temperature dependance is 
%The data in the apparently standard regime is examined by utilizing 
%this model function in \S5.3.
$T(r)=T_{\rm in}\cdot (r/r_{\rm in})^{-p}$. This 
$p$-free disk model differs from the standard MCD model which has
$p$ fixed at $3/4$, so there are now 
three fit parameters, $T_{\rm in}$, $r_{\rm in}$, and $p$ 
(see also Hirano et al.1995, Mineshige et al. 1994).

\begin{figure}[htbp]
%fig7a fig7b fig7c
\begin{center}
\begin{minipage}{0.32\textwidth}
\footnotesize{(a) LMC X-3}

\vspace*{-3mm}
\includegraphics[clip=true,width=3cm,angle=-90]
{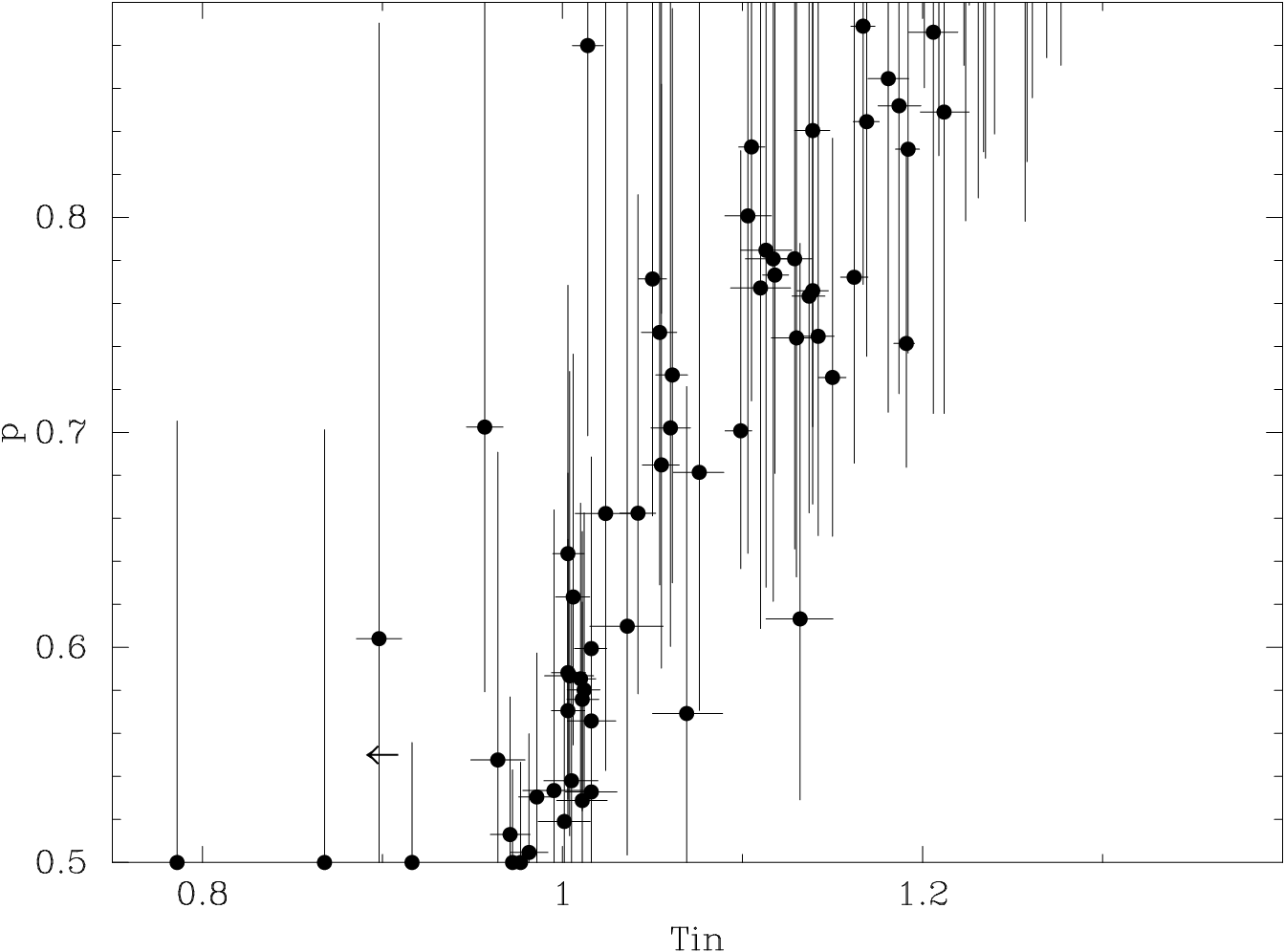}
\end{minipage}
\begin{minipage}{0.32\textwidth}
\footnotesize{(b) GRO J$1655-40$}

\vspace*{-3mm}
\includegraphics[clip=true,width=3cm,angle=-90]
{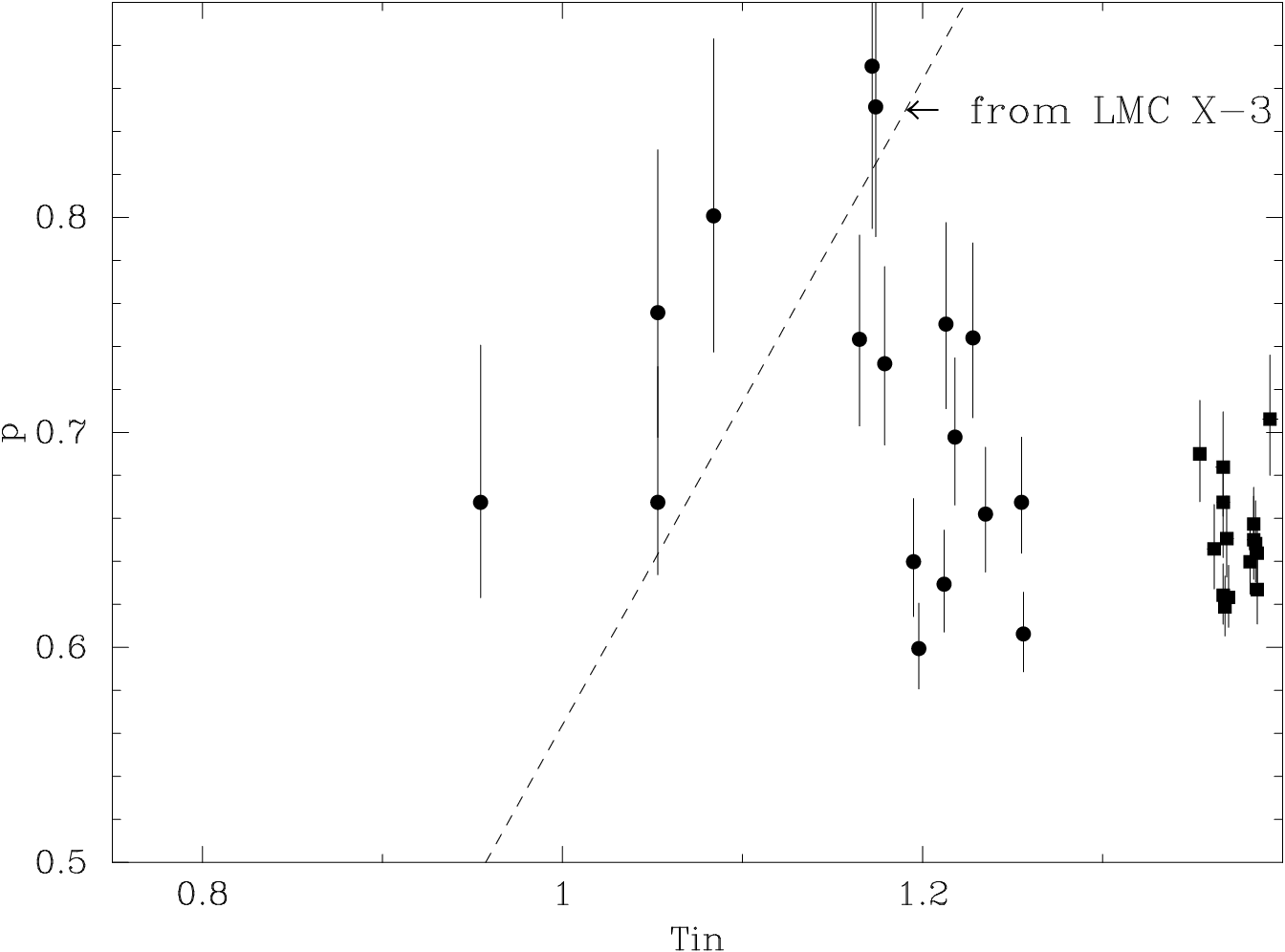}
\end{minipage}
\begin{minipage}{0.32\textwidth}
\footnotesize{(c) XTE J$1550-564$}

\vspace*{-3mm}
\includegraphics[clip=true,width=3cm,angle=-90]
{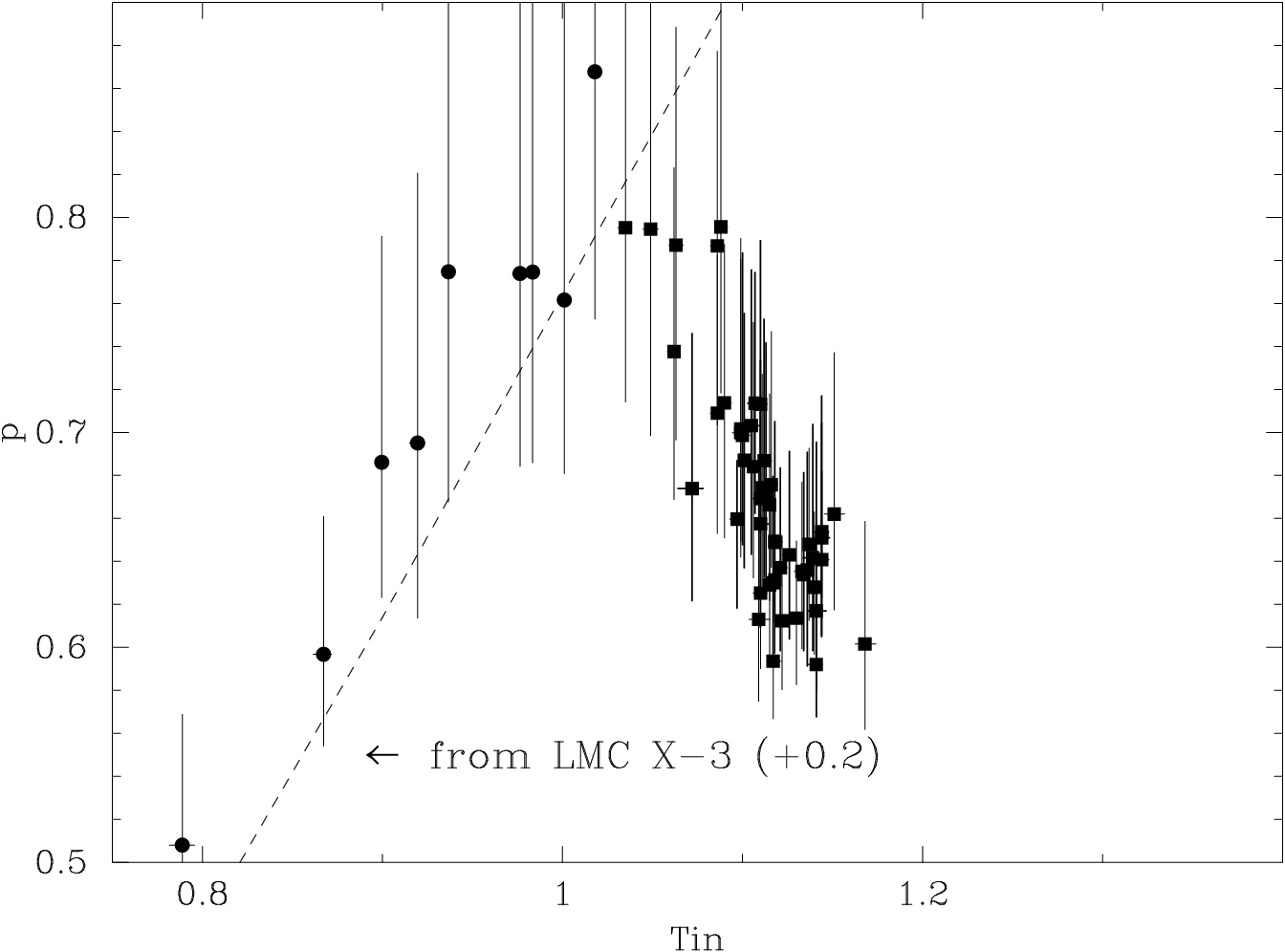}
\end{minipage}
\end{center}
\caption{The best fit values of $p$ against $T_{\rm in}$
of LMC X-3 (a), GRO J$1655-40$ (b), and XTE J$1550-564$ (c). 
Instead of the values of $T_{\rm in}$ obtained
 by the $p$-free disk model, those by the MCD model are
employed, in order to avoid any systematic coupling
between $p$ and $T_{\rm in}$.
Dashed lines show the calibrated trace of $p$ obtained with the LMC X-3 (a). 
The absolute value of $p$ includes PCA systematic effects, and the calibrated line in panel (c) is obtained by shifting the result of LMC X-3 as $\Delta p=+0.2$.}
\label{fig:t-p}
\end{figure}

In Fig.~\ref{fig:t-p}, the best fit values of $p$ are plotted against $T_{\rm in}$ for 
LMC~X-3, GRO~J$1655-40$ and XTE~J$1550-564$.
Though the absolute values of $p$ were slightly affected by fitting conditions 
(whether $\Gamma$ or $N_{\rm H}$ were fixed or free, 
including (or excluding) a gaussian for a narrow iron line), 
the characteristic $p$-$T_{\rm in}$ behavior on Fig.~\ref{fig:t-p} 
was not affected by them.
As a calibration trace on $p$-$T_{\rm in}$ plane for the 
standard regime, we show a result of LMC~X-3 in Fig.~\ref{fig:t-p}a. 
This source stays in the standard regime up to $\sim L_{\rm E}$, 
and shows positive correlation between $p$ and $T_{\rm in}$.  
With the PCA data, 
this positive correlation is considered as an artifact which is present 
even if the standard accretion disk is realized. 
This artifact happens 
because the MCD model ignores the boundary condition at the inner 
edge of the standard disk solution. 
The actual temperature gradient of a standard accretion disk must 
be flatter than 3/4 near the innermost disk edge, where
the temperature will approach zero. 
As $T_{\rm in}$ decreases, 
the limited PCA band pass will sample
preferentially the emission from inner disk regions, thus making $p$ 
appear smaller than 3/4 as is found in LMC~X-3.

As for XTE~J$1550-564$, 
while the data points for $T_{\rm in}\le1$keV (standard regime)
are well consistent with the calibrated trace in Fig.~\ref{fig:t-p}c, 
those for $T_{\rm in}\ge 1$~keV (in the apparently standard regime) 
appear well below the trace 
(and in fact, $p$ {\sl negatively} correlates to $T_{\rm in}$). 
Similarly, the data points of GRO~J$1655-40$ for $T_{\rm in}\le 1$--1.1~keV 
are mostly consistent with the trace, but those for $T_{\rm in}\ge1.1$~keV 
are below the trace.
Therefore, the observed temperature gradient in the apparently standard regime 
is smaller than in the standard regime. 

This result means that the radiative efficiency at the inner portion of the 
accretion disk in the apparently standard regime is lower than that in the 
standard regime (case of the standard disk). 
Hence the accretion disk in the apparently standard regime 
deviates from the standard disk and requires another cooling process in addition to 
radiative cooling.
The advection in the optically thick disk 
is a candidate for the cooling process, and in fact, the slim disk solution, 
seen in a top slope of the $S$-shape sequence (Abramowicz et al. 1995), 
predicts smaller values of $p$ than the standard disk (e.g., Watarai et al. 2000). 

\subsection{Brief summary of the obtained picture}

We show that there are three spectral regimes in the stable high/soft state 
black hole binaries, the standard regime, the anomalous (weak very high) regime, 
and the apparently standard regime. The last two regimes appear typically when 
the X-ray luminosity exceeds a certain critical luminosity. The accretion disk 
structures are somewhat different in these three regimes, but all are consistent 
with the disk extending down to a constant inner radium, most probably the 
innermost stable Keplerian orbit around the central black hole. The accretion 
disk in each regime is characterized by a radiation dominated standard disk but 
in the anomolous regime there is also high energy plasma which causes the 
observed strong Compton scattering, while in the apparently standard regime 
there is some change to the cooling process, perhaps marking the 
onset of advection cooling in addition to 
radiative cooling. This picture can also explain many other black hole binaries in 
the high accretion rate state including 4U~$1630-47$ (Abe et al. 2004).

\section{Analogy to ULXs}

Based on their X-ray luminosities, 
ultraluminous compact X-ray sources (ULXs) in nearby galaxies are supposed to be 
intermediate mass ($\sim$30--100~$M_\odot$) black holes (Makishima et al.~2000),
super Eddington stellar 
mass black holes, or (mild) beaming black holes (e.g., King et al. 2001).
Among many ULXs, Mizuno et al. (2001) analyzed three bright 
(0.2--1.5$\times10^{40}~{\rm erg~s^{-1}}$) variable ULXs observed with ASCA, 
IC~342 source~1, M81~X-6 (sometimes called X-11) and NGC~1313 source~B. 
They found that the X-ray spectra are well reproduced by the MCD model 
and that obtained values of $R_{\rm in}$ are not constant but variable as 
$R_{\rm in}\propto T_{\rm in}^{-1}$. 
%(i.e., {\sl not} $L_{\rm disk}\propto T_{\rm in}^4$ but  $L_{\rm disk}\propto T_{\rm in}^2$. 
This is very similar to that of XTE~J$1550-564$ in the apparently standard regime.
In addition, Kubota et al. (2002) analyzed the other spectral data of IC~342 
source 1 obtained with ASCA in different date. 
During that observation, the source showed a power-law like spectrum rather than the 
MCD shape, and it was reanalyzed in the framework of the 
inverse comptonization as is the case of the anomalous regime. 
They found that this power-law like spectrum
then lies nicely on the same luminosity-temperature relation as defined 
by the MCD-type spectra given by Mizuno et al. (2001). 

We can see that the data points of bright ULXs on 
$T_{\rm in}$-$L_{\rm disk}$ diagram is understood by shifting up those of 
stellar black hole, XTE~J$1550-564$ in the brightest regimes 
(see Fig.~2 by Kubota et al. 2002).
These spectral studies thus supports 
the scenario of high accreting relatively massive (more massive than 
Galactic stellar black holes) black holes for ULXs.

 \vspace*{3mm}
%\section*{Acknowledgements}
We would like to thank C. Done, 
T. Mizuno, K. Ebisawa, K. Watarai, and S. Mineshige for 
helpful discussions. 
A.K. is supported by special postdoctoral researchers program in RIKEN.

%\vspace*{-5mm}

\end{document}